\documentclass[preprint,12pt]{elsarticle}

\usepackage{amsmath}
\usepackage{amssymb}
\usepackage{xcolor}
\usepackage{natbib}
\usepackage{tikz}
\usetikzlibrary{arrows.meta, positioning, calc, shapes.geometric}
\usepackage{caption}
\usepackage{multirow}
\usepackage{graphicx}
\usepackage{listings}
\usepackage{tcolorbox}
\tcbset{colback=white, colframe=black!30, boxrule=0.6pt,arc=3pt}

\newcommand{\abd}[1]{}
\newcommand{\ar}[1]{}

\biboptions{authoryear}

\usepackage[bookmarks=false]{hyperref}
    \hypersetup{colorlinks,
      linkcolor=blue,
      citecolor=blue,
      urlcolor=blue}

\begin{document}
\begin{frontmatter}

\title{Early Pruning for Public Transport Routing\tnoteref{wctr}}
\tnotetext[wctr]{Accepted for presentation at the World Conference on Transport Research (WCTR) 2026, Toulouse. To appear in Transportation Research Procedia.}

\author[a]{Andrii Rohovyi\corref{cor1}} 
\author[b]{Abdallah Abuaisha}
\author[a]{Toby Walsh}

\cortext[cor1]{Corresponding author. Tel.: +38-093-910-67-66}

\address[a]{Department of Computer Science and Engineering, University of New South Wales (UNSW), Sydney, NSW 2033, Australia}
\address[b]{Department of Data Science and Artificial Intelligence, Monash University, Melbourne, Australia}

\begin{abstract}

Routing algorithms for public transport, particularly the widely used RAPTOR and its variants, often face performance bottlenecks during the transfer relaxation phase, especially on dense transfer graphs, when supporting unlimited transfers. This inefficiency arises from iterating over many potential inter-stop connections (walks, bikes, e-scooters, etc.). To maintain acceptable performance, practitioners often limit transfer distances or exclude certain transfer options, which can reduce path optimality and restrict the multimodal options presented to travellers.

This paper introduces Early Pruning, a low-overhead technique that accelerates routing algorithms without compromising optimality. By pre-sorting transfer connections by duration and applying a pruning rule within the transfer loop, the method discards longer transfers at a stop once they cannot yield an earlier arrival than the current best solution.

Early Pruning can be integrated with minimal changes to existing codebases and requires only a one-time preprocessing step. The technique preserves Pareto-optimality in extended-criteria settings whenever the additional optimization criteria are monotonically non-decreasing in transfer duration. Across multiple state-of-the-art RAPTOR-based solutions, including RAPTOR, ULTRA-RAPTOR, McRAPTOR, BM-RAPTOR,  ULTRA-McRAPTOR, and UBM-RAPTOR and tested on the Switzerland and London transit networks, we achieved query time reductions of up to 57\%. This approach provides a generalizable improvement to the efficiency of transit pathfinding algorithms.

Beyond algorithmic performance, Early Pruning has practical implications for transport planning. By reducing computational costs, it enables transit agencies to expand transfer radii and incorporate additional mobility modes into journey planners without requiring extra server infrastructure. This is particularly relevant for passengers in areas with sparse direct transit coverage, such as outer suburbs and smaller towns, where richer multimodal routing can reveal viable alternatives to private car use.

\end{abstract}

\begin{keyword}
Pathfinding \sep Public Transit Routing \sep Multimodal Journey Planning \sep RAPTOR \sep Transport Policy \sep Mobility-as-a-Service
\end{keyword}

\end{frontmatter}

\section{Introduction}

Public transport journey planners are critical infrastructure for modern transit systems. They serve as the primary interface through which passengers discover and evaluate travel options, and their quality directly influences mode choice decisions. When a journey planner returns slow, incomplete, or suboptimal results, for example, by omitting feasible walking or cycling connections between stops, then travellers may perceive public transport as inconvenient and opt for private vehicles instead. Conversely, responsive planners that present rich multimodal options can increase transit ridership by revealing attractive journey alternatives that passengers would not otherwise discover. The computational engines behind these planners must balance two competing demands. On the one hand, transit agencies increasingly want to incorporate diverse transfer modes: walking, cycling, e-scooters, and other shared mobility services to support Mobility-as-a-Service (MaaS) strategies and broader sustainability goals. On the other hand, the resulting transfer graphs become denser and more computationally expensive to traverse, threatening the sub-second response times that interactive applications require.

To address this tension, we introduce Early Pruning, a technique that accelerates the transfer relaxation phase of RAPTOR-based routing algorithms without compromising optimality. The core idea is to pre-sort all outgoing transfer edges at each stop by their duration, then terminate the iteration as soon as the next transfer cannot yield an earlier arrival than the current best solution at the target. Because the edges are sorted, all subsequent transfers are guaranteed to be at least as long, and can be safely skipped. This simple rule requires only a one-time preprocessing step (edge sorting) that takes under 600\,ms on tested networks and does not need to be recomputed when timetable schedules change. Across six RAPTOR variants and two real-world transit networks, Early Pruning achieves query time reductions of up to 57\%.

RAPTOR \cite{raptor} is the most widely deployed transit routing algorithm, powering Bing Maps, OpenTripPlanner, R5, Navitia.io, Solari, and many other systems. Its round-based design naturally supports multi-criteria search and parallelisation, making it attractive for production use. However, RAPTOR and its variants all share a common bottleneck: the transfer relaxation phase, where the algorithm iterates over outgoing transfer edges from each updated stop. When the transfer graph is dense---as it becomes when incorporating walking, cycling, and micro-mobility connections, or when using the transitive closure required by some variants---this phase dominates the total query time. Early Pruning is effective on both the original dense graphs and on ULTRA shortcuts \cite{baum2019ultra}, and remains valid across timetable updates.

In our work, we propose a method, which we call Early Pruning, and prove its correctness for both single-criterion and extended-criterion settings. We evaluate it across six RAPTOR variants on two real-world transit networks (Switzerland and London), demonstrating consistent speedups of 2--57\%. Also, we analyse the relationship between graph density and performance improvement. Finally, we discuss the implications for transport planning, arguing that faster routing algorithms can support broader policy objectives, including multimodal integration, equitable access, and Mobility-as-a-Service adoption.

\section{Literature Review}

This section positions Early Pruning within the broader landscape of transit routing research. The algorithms themselves are formally defined in Section~3.

The development of efficient algorithms for public transit routing has been a sustained research effort spanning over two decades. Early approaches adapted Dijkstra's algorithm \cite{dijkstra1959note} and its extensions to the transit network \cite{multi_dijkstra, rohovyyi2024ttn} for time-dependent or time-expanded graph models, but the resulting representations were often large and inefficient for the temporal structure of timetable data. These limitations motivated algorithms that operate directly on timetable data, including the Connection Scan Algorithm (CSA) \cite{csa}, Trip-Based Routing \cite{trip_based_routing, trip_based_routing_condense}, Transfer Patterns \cite{transfer_patterns}, and TCD \cite{abuaisha2024efficient}. Among these, RAPTOR \cite{raptor} has achieved the broadest practical adoption due to its round-based design, which naturally produces Pareto-optimal journeys. The RAPTOR framework has since been extended in many directions: MR-$\infty$ (MR) and MCR \cite{delling2013computing} enable unlimited transfers via Dijkstra-based transfer relaxation; BM-RAPTOR \cite{delling2019fast} and UBM-RAPTOR \cite{sauer2024closing} constrain the Pareto space for faster queries with additional criteria; and the ULTRA framework \cite{baum2019ultra} precomputes shortcut edges to avoid transitive closure entirely. Wagner and Z{\"u}ndorf \cite{unrestricted_walking} demonstrated the significant impact of unrestricted walking on travel times using a profile algorithm based on MCR. Phan and Viennot \cite{phan2019hub} subsequently explored hub labeling as an alternative approach, decomposing walking transfers into two consecutive hops and achieving significant speedups for both RAPTOR and CSA-based algorithms. Dijkstra-based approaches have also continued to evolve in parallel, with recent work addressing unlimited-transfer routing \cite{rohovyi2025multimodal} 

Pruning strategies are well-established in the transit routing literature. The original RAPTOR paper \cite{raptor} introduced several forms of pruning: \emph{marking} (only processing routes that serve recently updated stops), \emph{local pruning} (discarding labels dominated at a stop), and \emph{target pruning} (discarding labels dominated by the best known arrival at the target). Our Early Pruning technique is most closely related to target pruning, but operates at a different granularity. While standard target pruning checks whether a \emph{label} at a stop is dominated by the target's best known arrival, Early Pruning exploits the \emph{ordering} of transfer edges to terminate the entire transfer iteration at a stop early. This is possible because pre-sorting transfers by duration guarantees that all subsequent transfers will produce later arrivals than the current one. The same principle was introduced by \cite{phan2019hub} in the context of hub labellings, where sorted hub lists enable a target-pruning optimisation that stops scanning once arrival times exceed the best-known target time. Our work applies this principle directly within the transfer relaxation loop of RAPTOR, without requiring hub label preprocessing, and extends it to settings with additional criteria, such as walking duration, including McRAPTOR, BM-RAPTOR, and their ULTRA-based counterparts.

The algorithmic advances described above have significant implications for transport planning and policy. Modern transport policy increasingly emphasises multimodal integration, where public transit serves as the backbone of a broader mobility ecosystem that includes walking, cycling, ride-sharing, and micro-mobility services. The Mobility-as-a-Service (MaaS) paradigm aims to provide seamless, door-to-door journeys by integrating these modes into a single planning and payment platform \cite{kamargianni2016maas, jittrapirom2017maas}. Realising this vision requires journey planners that can efficiently evaluate multimodal transfer options in real time. When routing algorithms are too slow to process dense, multimodal transfer graphs, practitioners must restrict transfer distances or exclude modes, limiting the journey options available to passengers. This creates a direct link between algorithmic performance and the quality of the transport service experienced by users.

\section{Preliminaries}

This section defines the network model, the key algorithms used in our experiments, and the notation referenced throughout the paper.

\subsection{Terminology}

A public transit network is formally defined as a tuple $(\Pi, S, T, R, G)$. Here, $\Pi \subseteq N_0$ represents the set of all possible time units within a period of operation, such as the seconds of a day. $S$ is a finite set of stops, each corresponding to a distinct location where passengers can board or alight a vehicle, like a bus stop or train platform. $T$ is a finite set of trips, where each trip $t \in T$ denotes a specific vehicle's journey along a line, visiting a sequence of stops; at each stop $p$ in the sequence, the vehicle may drop off or pick up passengers. Additionally, $R$ is a finite set of routes, and the trips in $T$ are partitioned into these routes, with each route $r \in R$ consisting of all trips that follow the same sequence of stops. It is typical for there to be many more trips than routes.

$G$ is a weighted transfer graph, defined as a tuple $(V, E)$. $V$ is a set of vertices where $S \subseteq V$, and $E$ is a set of edges, where $E \subseteq V \times V$. For each edge $e = (u, v) \in E$, there is a transfer time $\tau(u, v)$. $G$ is called \textit{transitively closed} if for every pair of vertices $u, v \in V$ with an indirect path between them, there exists a direct edge $(u, v) \in E$. The \emph{density} of $G$, denoted as $\rho(G)$, is defined as the ratio of the number of edges in $G$ to the maximum possible number of edges. For the directed graph it is equal to: $\rho(G) = \frac{|E|}{|V|(|V|-1)}$. The density $\rho(G)$ takes values in the interval $[0,1]$, where $0$ corresponds to a graph with no edges and $1$ corresponds to a complete graph. The path is the sequence of stops in transit network $P = \langle v_1, \dots, v_k \rangle$, where $\tau(P) := \sum_{i=1}^{k-1} \tau(v_i, v_{i+1})$.

A journey $J$ describes the movement of a passenger through the network from a source vertex $s \in V$ to a target vertex $t \in V$. Formally, a journey is a sequence $J = \langle x_1, x_2, \dots, x_n \rangle$, where each element $x_i$ is either a trip $t \in T$ or a transfer $(u,v) \in E$, taken in the order of travel. The journey satisfies the connectivity condition that the end stop of $x_i$ coincides with the start stop of $x_{i+1}$ for all $i = 1, \dots, n-1$.

\subsection{Algorithms}
Dijkstra's algorithm \cite{dijkstra1959note} is a greedy algorithm for finding the shortest paths in a weighted graph with non-negative edge weights. It starts at a source node and tracks nodes with known shortest-path distances. In each step, it picks the unvisited node with the smallest known distance from the source and relaxes its outgoing edges, updating the distances of nearby nodes if a shorter path is found. The algorithm ends when all reachable nodes are visited, ensuring the shortest path to each node from the source.

RAPTOR \cite{raptor} is a round-based pathfinding algorithm for public transport networks. Its main idea is that the first public transport option you can reach often provides a better arrival time than those accessed later. It works in rounds: in each round $i$, it manages all relevant journeys with exactly $i-1$ transfers. This approach also allows you to optimise two factors simultaneously: the earliest arrival time and the number of transfers. In each round, RAPTOR iterates over transfers (walk, e-scooter, etc.) in the transfer graph departing from stops updated in the previous round, updating the arrival times accordingly.

McRAPTOR \cite{raptor} extends RAPTOR to optimise additional criteria beyond arrival time and number of transfers, such as walking duration. Given a set of criteria, a journey $J$ \emph{dominates} another journey $J'$, if $J$ is not worse than $J'$ in any criterion. In our research, we use a three-criteria version of McRAPTOR, which includes walking duration, arrival time, and the number of transfers. Although McRAPTOR can be very useful in practice, as it offers users various options that can be sorted and selected according to their preferences, it has a long query time. One way to speed it up is to find the solution not within the entire search space, but within a defined, restricted Pareto space. This modification, called BM-RAPTOR \cite{delling2019fast}, reduces the query time by 65 times compared to McRAPTOR, which is still roughly twice as slow as standard RAPTOR \cite{sauer2024closing}.

ULTRA \cite{baum2019ultra} is a precomputation technique that calculates the Pareto optimal set of shortcuts in the transfer graph, significantly speeding up future searches. The modified version of RAPTOR that operates on ULTRA shortcuts is called ULTRA-RAPTOR; it allows unlimited transfer queries at speeds comparable to traditional RAPTOR. The extended-criteria modification is called ULTRA-McRAPTOR, and its bounded version is named UBM-RAPTOR \cite{sauer2024closing}.

\section{Early Pruning}

To speed up the algorithms described above, we use a simple pruning rule. The core idea is to stop iterating over transfers as soon as they become non-competitive with the currently known best path to the target. To do this, we presort all outgoing edges in $G$ by their duration, which ensures that later edges in the iteration are at least as long as earlier ones. As a result, we can terminate the loop over outgoing edges if the arrival time at the next stop already equals or exceeds the best known arrival time at the target.

Let $s$ be a stop updated in the current round, with arrival time $\tau_k(s)$, and let $t$ be the query target with current best arrival time $\tau^*(t)$. Let the outgoing transfer edges from $s$ be sorted in non-decreasing order of their duration: $e_1, e_2, \dots, e_m$ where $\tau(e_1) \leq \tau(e_2) \leq \dots \leq \tau(e_m)$. For each edge $e_i = (s, v_i)$, the arrival time at $v_i$ via this transfer is $\tau_k(s) + \tau(e_i)$. If $\tau_k(s) + \tau(e_i) \geq \tau^*(t)$, then for all $j > i$:
\[
\tau_k(s) + \tau(e_j) \geq \tau_k(s) + \tau(e_i) \geq \tau^*(t)
\]
Therefore, no transfer $e_j$ with $j > i$ can improve the arrival at $t$, and the iteration can be terminated.

\begin{figure}[ht]
\centering
\resizebox{0.95\textwidth}{!}{%
\begin{tikzpicture}[
  stop/.style={circle,draw,inner sep=2pt,minimum size=20pt},
  target/.style={rectangle,draw,rounded corners,inner sep=3pt,minimum size=20pt},
  pruned/.style={->, red, thick, dashed, opacity=0.6},
  >=Stealth
]

\node[stop] (s_src) {s};
\node[stop,right=2cm of s_src] (u) {u};
\node[stop,right=3cm of u,yshift=2.5cm] (a) {a};
\node[stop,right=3cm of u]  (b) {b};
\node[stop,right=3cm of u,yshift=-2.5cm] (c) {c};
\node[target,right=2.5cm of b] (t) {t};

\draw[->, dashed] (s_src) -- (u);
\node[above=0.15cm of u, anchor=south east, font=\small] {arrives 1:55\,p.m.};

\draw[->, thick] (u) -- node[above,sloped,font=\small,pos=0.45] {2 min} (a);
\draw[pruned] (u) -- node[below,font=\small,pos=0.3] {5 min} (b);
\draw[pruned] (u) -- node[below,sloped,font=\small,pos=0.45] {9 min} (c);

\draw[->,dashed] (a) -- (t);
\draw[->,dashed] (b) -- (t);
\draw[->,dashed] (c) -- (t);

\node[red, font=\Large\bfseries] at ($(u)!0.6!(b)$) {\texttimes};
\node[red, font=\Large\bfseries] at ($(u)!0.6!(c)$) {\texttimes};

\node[red, font=\small\bfseries, anchor=south] at ($(u)!0.55!(b) + (0.3,0.3)$) {prune here};

\node[right=0.6cm of t, align=left, font=\small] {
  best known arrival \\ at target = 2:00\,p.m.
};

\end{tikzpicture}%
}
\caption{Early Pruning: transfers from stop $u$ are pre-sorted by duration. Once the arrival time via a transfer equals or exceeds the best known arrival at $t$, all remaining transfers are pruned.}\label{tab:early_prun}
\end{figure}
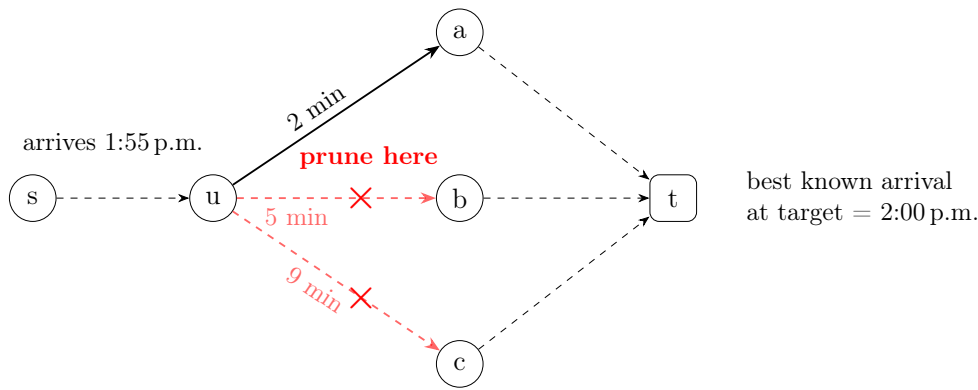

An example of Early Pruning is shown in Figure~\ref{tab:early_prun}, where a passenger arrives at stop $u$ at 1:55 p.m. The transfers from $u$ to other stops ($a, b, c$) are pre-sorted by duration (2 min, 5 min, 9 min). The algorithm iteratively checks outgoing transfer edges in ascending order, starting from the $u \to a$ transfer. Upon checking the next transfer, $u \to b$, the passenger reaches $b$ at 2:00 p.m. ($1{:}55 \text{ p.m.} + 5 \text{ min}$). If the current best arrival time at the target $t$ is already 2:00 p.m., checking $u \to b$ and all subsequent, longer transfers (like $u \to c$) is futile, as they cannot result in an earlier arrival at $t$. Thus, the method \emph{prunes} these longer transfers (marked with red dashed lines and `$\times$') and skips the associated computation, leading to speedups, especially on dense transfer graphs.

The same idea extends to extended-criteria variants such as McRAPTOR, ULTRA-McRAPTOR, BM-RAPTOR, and UBM-RAPTOR, provided a monotonicity condition holds: each additional criterion beyond arrival time and number of transfers must be monotonically non-decreasing in the transfer duration. Under this condition, once a partial path from stop $u$ is dominated in the Pareto sense by an existing solution to target $t$, all subsequent transfers (which are longer, due to pre-sorting) will produce paths that are at least as bad in every criterion. Thus they can be safely pruned. 

\textbf{Remark 1.} \textit{Let $s$ be a stop with arrival time $\tau_k(s)$, and let $e_i = (s, v_i)$ and $e_j = (s, v_j)$ be two outgoing transfers with $\tau(e_i) \leq \tau(e_j)$. Early Pruning preserves Pareto-optimality if for every optimization criterion $c \in C$, the label produced by taking transfer $e_j$ is not better than the label produced by taking transfer $e_i$. That is, a longer transfer from the same stop cannot improve any criterion.}

In our experiments, we use three criteria: arrival time, walking duration, and number of transfers. A longer transfer from the same stop leads to a later arrival time and a longer walk, while the number of transfers remains unchanged. All three criteria therefore satisfy the monotonicity condition from Remark~1. The same holds for other transfer mode durations, such as cycling time or e-scooter travel time, making Early Pruning applicable to extended-criteria formulations beyond the specific configuration tested here.

We also test this idea on the CSA \cite{csa} algorithm. However, despite the algorithmic improvement, it does not improve results in practice. This is because CSA uses a smaller cache in memory; when it scans connections and loads needed information, the pruning remains unchanged. In contrast, RAPTOR has a much larger cache, which is why the speed-up is visible. The valuable comment related to this was left by one of the ULTRA package contributors to our PR.\footnote{\url{https://github.com/kit-algo/ULTRA/pull/31\#issuecomment-3170502692}}

\section{Experiments}

For experiment and test purposes, we modify the original implementation of all of the mentioned algorithms from the Karlsruhe Institute of Technology.\footnote{\url{https://github.com/kit-algo/ULTRA}} We created a fork of this repository and made it publicly available.\footnote{\url{https://github.com/andrii-rohovyi/PublicTransitRoutingWithUnlimitedTransfer}}

The computations and analyses were conducted using the GNU C++ Compiler (g++) version 13.3.0. The experiments were performed on a MacBook Pro with an Apple M3 Max chip, featuring a 16-core CPU, a 40-core GPU, and 128 GB of unified memory. The Colima container runtime was utilised, configured with the x86\_64 architecture, 16 CPU cores, 128 GB of memory, and a 500 GB disk.

The experiments were conducted on the London and Switzerland networks, which were previously accessible from the Karlsruhe Institute of Technology (KIT).\footnote{\url{https://i11www.iti.kit.edu/PublicTransitData/ULTRA/}}. KIT \cite{baum2019ultra} constructed Unrestricted Transfer Graphs by extracting road graphs, including pedestrian zones and staircases, from OpenStreetMap\footnote{\url{https://download.geofabrik.de/}}. Unless stated otherwise, they used walking as the transfer mode, assuming a constant speed of 4.5 km/h. However, the Unrestricted Transfer Graph cannot be used with algorithms that require a transitively closed graph, since the transitive closure would be too large \cite{unrestricted_walking}. To address this, the authors precomputed more compact Transitive Transfer Graphs by adding edges between all stops whose transfer-graph distance was below a chosen threshold (9 minutes for Switzerland and 4 minutes for London) and then computing the transitive closure. The thresholds were set so that the resulting graph had an average vertex degree of about 100 \cite{baum2019ultra}. Finally, two types of ULTRA precomputations were performed on the Unrestricted Transfer Graphs to generate shortcuts: one for earliest-arrival algorithms and another for extended-criteria algorithms. Networks statistics used in the experiments are provided in Table \ref{tab:network}.

\begin{table}[h]
    \centering
    \caption{Networks Statistics \abd{I suggest to consider making font size in the table smaller to be consistent with Table 2 (change size to footnotesize?)} \ar{Good catch fixed}}\label{tab:network}
    \footnotesize
    \setlength{\tabcolsep}{5pt}
    \begin{tabular}{|l|r|r|}
        \hline
        \textbf{Metric} & \textbf{Switzerland} & \textbf{London} \\
        \hline
        \multicolumn{3}{|c|}{\textbf{Transport Network Info}} \\
        \hline
        Stops & 25,125 & 19,682 \\
        Routes & 13,786 & 1,955 \\
        Trips & 350,006 & 114,508 \\
        Stop events & 4,686,865 & 4,508,644 \\
        \hline
        \multicolumn{3}{|c|}{\textbf{Transfer Graph (Transitive)}} \\
        \hline
        Vertices & 25,125 & 19,682 \\
        Edges & 3,212,206 & 2,639,402 \\
        Density & $0.88 \times 10^{-5}$ & $8.00 \times 10^{-5}$ \\
        Edges sorting time & 417 ms 889 $\mu$s & 567 ms 746 $\mu$s \\
        \hline
        \multicolumn{3}{|c|}{\textbf{ULTRA Shortcuts (Shortcuts)}} \\
        \hline
        Number of shortcuts & 170,713 & 190,388 \\
        Density & $0.047 \times 10^{-5}$ & $0.58 \times 10^{-5}$ \\
        Edges sorting time & 24 ms 863 $\mu$s & 24 ms 189 $\mu$s \\
        \hline
        \multicolumn{3}{|c|}{\textbf{ULTRA Multicriteria Shortcuts (Mc Shortcuts)}} \\
        \hline
        Number of shortcuts & 218,637 & 135,379 \\
        Density & $ 0.06 \times 10^{-5}$ & $ 0.411\times 10^{-5}$ \\
        Edges sorting time &30 ms 874 $\mu$ &20 ms 200 $\mu$ \\
        \hline
    \end{tabular}
\end{table}

Presorting the edges takes negligible time, as shown in Table~\ref{tab:network}. It is a quick one-time operation that is delay-robust and does not need to be recomputed if the schedule changes, unlike the ULTRA algorithm. The sorting time for the London transitive graph is slightly higher than for Switzerland despite having fewer edges. This is attributable to London's high graph density ($8.00 \times 10^{-5}$ vs \ $0.88 \times 10^{-5}$), which results in more edges per vertex and less favourable cache access patterns during sorting.
\abd{why is edge sorting time in Transitive Graph (Table 1) longer for London despite having fewer edges? add comment/explanation in text?} \ar{Added, but tbh I am also not sure what caused it.}

\begin{table}[h]
    \centering
    \caption{Query time}\label{tab:run}
    \resizebox{\textwidth}{!}{%
    \begin{tabular}{|l|l|r|r|r|r|r|r|}
        \hline
        \multirow{2}{*}{\textbf{Algorithm}} & \multirow{2}{*}{\textbf{Graph}} & \multicolumn{3}{c|}{\textbf{Switzerland}} & \multicolumn{3}{c|}{\textbf{London}} \\
        \cline{3-8}
        && Original & Early Pruning & Speedup & Original & Early Pruning & Speedup \\
        \hline
        RAPTOR & Transitive & 53.8\,ms & \textbf{42.1\,ms} & 21.7\% & 48.6\,ms & \textbf{32.2\,ms} & 33.9\% \\
        ULTRA-RAPTOR & Shortcuts & 44.3\,ms & \textbf{39.4\,ms} & 11.1\% & 26.3\,ms & \textbf{22.7\,ms} & 13.6\% \\
        McRAPTOR & Transitive & 2997\,ms & \textbf{1646\,ms} & 45.1\% & 3966\,ms & \textbf{1707\,ms} & 56.9\% \\
        ULTRA-McRAPTOR & Mc Shortcuts & 749\,ms & \textbf{731\,ms} & 2.4\% & 477\,ms & \textbf{462\,ms} & 3.1\% \\
        BM-RAPTOR & Transitive & 196\,ms & \textbf{183\,ms} & 6.6\% & 258\,ms & \textbf{217\,ms} & 16.0\% \\
        UBM-RAPTOR & Mc Shortcuts & 118\,ms & \textbf{116\,ms} & 1.6\% & 82\,ms & \textbf{77\,ms} & 6.1\% \\
        \hline
    \end{tabular}%
    }
\end{table}

Query performance of algorithms was averaged over 1,000 random queries. BM-RAPTOR and UBM-RAPTOR have been run with arrival slack and trip slack equal to 1.25, as suggested in \cite{sauer2024closing}. All query time results can be found in Table~\ref{tab:run}. The results clearly show that Early Pruning consistently reduces query times across all tested networks. The most pronounced improvements are observed for methods such as RAPTOR, McRAPTOR, and BM-RAPTOR, because these algorithms operate on the transitive graph, which contains a large number of edges. Pruning a subset of these edges leads to a substantial reduction in computation time. In contrast, the impact of Early Pruning on ULTRA-based RAPTOR variants is smaller, as these methods prune precomputed shortcuts, which are far fewer in number. Overall, the Pearson correlation \cite{pearson1895} between graph density and the performance improvement achieved through Early Pruning is 0.62 (statistically significant, $p \approx 0.033$), indicating that the denser the graph on which the algorithm is executed, the greater the speedup observed.

The runtime reductions we report translate directly into operational benefits for real-world journey planners. For a metropolitan public-transport system that handles millions of trip queries per day, a 10--30\% decrease in average query time allows agencies to support more complex multimodal options, such as flexible walking transfers or shared-mobility connections, without increasing server capacity. This improves the responsiveness of passenger-facing apps, enables planners to expand transfer radii to capture more realistic travel choices, and lowers infrastructure and energy costs for operators. Algorithms that do not require additional preprocessing, unlike ULTRA-based methods, benefit more from Early Pruning. Consequently, this technique is especially effective for dynamic algorithms operating in changing environments.

\section{Implications for Transport Policy and Planning}

\abd{While adding this section is important and directly addressing the reviewer's comment, I feel it is too long. Maybe remove subsection headers and make each of the four paragraphs a little more concise?} \ar{done}
\abd{I think it now appears more dense. My earlier suggestion to remove the subsection headers may not have been ideal. Still, I feel the four paragraphs are a bit long? Perhaps they could be made slightly suggest so the overall section is not too long. However, I am also happy to keep it as is if you think the current version works better.} \ar{Fixed, what do you think about the new version?}

The algorithmic improvements presented in this paper have practical consequences that extend beyond computational performance. In this section, we discuss how faster transit routing algorithms can support broader objectives in transport policy, multimodal planning, and equitable access to mobility.

Incorporating additional transfer modes into journey planners, as envisioned by Mobility-as-a-Service (MaaS) platforms that integrate walking, cycling, e-scooters, and shared mobility into a single interface, increases the density of the transfer graph, which directly increases query times. In practice, transit agencies must choose between offering comprehensive multimodal options and maintaining fast query responses. Early Pruning relaxes this trade-off: the 34\% speedup achieved for RAPTOR on the London transitive graph could allow an agency to increase the distance threshold used to compute the transitive closure from 4~minutes to a longer duration while maintaining the same query time, thereby including a richer set of transfer connections between stops. For queries that optimise additional criteria beyond arrival time and number of transfers, such as walking duration, the 57\% speedup achieved for McRAPTOR means that a planner returning results in 4~seconds can now respond in under 2~seconds, bringing it within the range of interactive applications.

For transit agencies handling millions of queries per day, a 10--30\% reduction in average query time translates directly into lower server and energy costs, or the ability to serve more queries with existing infrastructure. This is particularly relevant for agencies in developing regions or smaller cities that may lack the resources for extensive server infrastructure. Early Pruning is also well-suited to dynamic planning environments because its preprocessing (edge sorting) does not depend on timetable data. Unlike ULTRA shortcuts, which must be recomputed when the timetable changes, edge sorting only needs updating when new transfer options are added to the graph, and this recomputation is relatively fast (under 600\,ms on the tested networks).

Faster routing can also contribute to more equitable transit access. Under tight computational budgets, the first options dropped are typically complex multimodal transfers, precisely the options that matter most for passengers in areas with sparse direct transit coverage. By enabling a wider variety of transfer options to be evaluated in real time, Early Pruning helps passengers in outer suburbs and smaller towns discover viable journeys that combine, for example, cycling connections to the nearest transit stop with onward public transport, supporting broader goals around reducing car dependency.

\section{Conclusion and Future Work}

In this work, we introduce a new technique, Early Pruning, which accelerates various pathfinding algorithms in public transit routing. We adapted it for several RAPTOR modifications and observed consistent improvements across all variants, with speedups ranging from 2\% to 57\%, depending on network characteristics and algorithm modifications. In particular, denser networks benefit more from this approach. By enabling a speedup of up to 57\% on highly complex queries (like McRAPTOR), which provide users with additional criteria such as walking duration, Early Pruning makes advanced multimodal planning computationally feasible for large-scale production systems like trip planners. This not only results in a better user experience but can also potentially influence mode choice by delivering richer, faster routing information.

From a transport policy perspective, the results demonstrate that algorithmic optimisations of this kind are not merely technical improvements, they have tangible implications for the quality and breadth of journey-planning services that transit agencies can offer. Faster routing enables wider transfer radii, richer multimodal options, and more responsive passenger-facing applications, all of which can support policy objectives related to MaaS adoption, sustainable mobility, and equitable access. Importantly, because Early Pruning depends only on the sorting of transfer edges and not on timetable data, it is inherently robust to schedule changes caused by delays, cancellations, or other real-time disruptions.

For future work, we plan to explore the application of our technique to other pathfinding algorithms that involve iterative searches over transfer sets. Our experiments used the official RAPTOR implementation from the Karlsruhe Institute of Technology. Although this implementation is not widely adopted in the open-source community, Early Pruning can be integrated into popular platforms such as OpenTripPlanner, R5, Navitia.io, and others. This integration could significantly accelerate RAPTOR-based routing, as this algorithm is the most recognised in the open-source community.

\bibliographystyle{elsarticle-harv}
\bibliography{xampl}

@InProceedings{unrestricted_walking,
  author =	{Wagner, Dorothea and Z\"{u}ndorf, Tobias},
  title =	{{Public Transit Routing with Unrestricted Walking}},
  booktitle =	{17th Workshop on Algorithmic Approaches for Transportation Modelling, Optimization, and Systems (ATMOS 2017)},
  pages =	{7:1--7:14},
  series =	{Open Access Series in Informatics (OASIcs)},
  year =	{2017},
  volume =	{59},
  editor =	{D'Angelo, Gianlorenzo and Dollevoet, Twan},
  publisher =	{Schloss Dagstuhl -- Leibniz-Zentrum f{\"u}r Informatik},
  address =	{Dagstuhl, Germany},
  doi =		{10.4230/OASIcs.ATMOS.2017.7}
}

@inproceedings{trip_based_routing,
  author    = {Sascha Witt},
  title = {Trip-Based Public Transit Routing},
  booktitle = {Algorithms --- ESA 2015},
  editor    = {Bansal, Nikhil and Finocchi, Irene},
  series    = {Lecture Notes in Computer Science},
  volume    = {9294},
  pages     = {1025--1036},
  publisher = {Springer},
  year      = {2015},
  doi       = {10.1007/978-3-662-48350-3_85},
  isbn      = {978-3-662-48350-3}
}

@InProceedings{trip_based_routing_condense,
  author    = {Witt, Sascha},
  title     = {Trip-Based Public Transit Routing Using Condensed Search Trees},
  booktitle = {16th Workshop on Algorithmic Approaches for Transportation Modelling, Optimization, and Systems (ATMOS 2016)},
  pages     = {10:1--10:12},
  volume    = {54},
  year      = {2016},
  publisher = {Schloss Dagstuhl -- Leibniz-Zentrum f{\"u}r Informatik},
  doi       = {10.4230/OASIcs.ATMOS.2016.10}
}

@inproceedings{transfer_patterns,
    author    = {H. Bast and E. Carlsson and A. Eigenwillig and R. Geisberger and C. Harrelson and V. Raychev and F. Viger},
  title     = {Fast Routing in Very Large Public Transportation Networks Using Transfer Patterns},
  booktitle = {ESA 2010},
  year      = {2010},
  doi       = {10.1007/978-3-642-15775-2_25},
volume    = {6346},
pages = {290–301}
}

@inproceedings{raptor,
  author    = {Daniel Delling and Thomas Pajor and Renato F. Werneck},
  title     = {{Round-Based Public Transit Routing}},
  booktitle = {Proceedings of the 14th Meeting on Algorithm Engineering and Experiments (ALENEX)},
  year      = {2012},
  pages     = {130--140},
  publisher = {SIAM},
  doi       = {10.1137/1.9781611972924.13},
  url       = {https://doi.org/10.1137/1.9781611972924.13}
}

@article{csa,
  author    = {Julian Dibbelt and Thomas Pajor and Ben Strasser and Dorothea Wagner},
  title     = {{Connection Scan Algorithm}},
  journal   = {ACM Journal of Experimental Algorithmics},
  volume    = {23},
  pages     = {1.7:1--1.7:56},
  year      = {2018},
  publisher = {ACM},
  doi       = {10.1145/3274661}
}

@article{dijkstra1959note,
  title={A Note on Two Problems in Connexion with Graphs},
  author={Dijkstra, Edsger W.},
  journal={Numerische Mathematik},
  volume={1},
  number={1},
  pages={269--271},
  year={1959},
  publisher={Springer}
}

@article{multi_dijkstra,
title = {Time-dependent Networks as Models to Achieve Fast Exact Time-table Queries},
journal = {Electronic Notes in Theoretical Computer Science},
volume = {92},
pages = {3-15},
year = {2004},
note = {Proceedings of ATMOS Workshop 2003},
issn = {1571-0661},
doi = {https://doi.org/10.1016/j.entcs.2003.12.019},
author = {Gerth {Stølting Brodal} and Riko Jacob}
}

@inproceedings{baum2019ultra,
  author    = {Baum, Moritz and Buchhold, Valentin and Sauer, Jonas and Wagner, Dorothea and Z\"{u}ndorf, Tobias},
  title     = {{UnLimited TRAnsfers for Multi-Modal Route Planning: An Efficient Solution}},
  booktitle = {Proceedings of the 27th Annual European Symposium on Algorithms (ESA'19)},
  series    = {Leibniz International Proceedings in Informatics (LIPIcs)},
  volume    = {144},
  pages     = {14:1--14:16},
  publisher = {Schloss Dagstuhl -- Leibniz-Zentrum f{\"u}r Informatik},
  year      = {2019},
  doi       = {10.4230/LIPIcs.ESA.2019.14}
}

@inproceedings{delling2013computing,
  title={Computing Multimodal Journeys in Practice},
  author={Delling, Daniel and Dibbelt, Julian and Pajor, Thomas and Wagner, Dorothea and Werneck, Renato F.},
  booktitle={Proceedings of the 12th International Symposium on Experimental Algorithms (SEA'13)},
  volume={7933},
  pages={260--271},
  year={2013},
  publisher={Springer},
  series={Lecture Notes in Computer Science (LNCS)},
  doi={10.1007/978-3-642-38527-8_24}
}

@inproceedings{delling2019fast,
  title={Fast and Exact Public Transit Routing with Restricted Pareto Sets},
  author={Delling, Daniel and Dibbelt, Julian and Pajor, Thomas},
  booktitle={Proceedings of the 21st Workshop on Algorithm Engineering and Experiments (ALENEX'19)},
  pages={54--65},
  year={2019},
  organization={Society for Industrial and Applied Mathematics (SIAM)},
  doi={10.1137/1.9781611975499.5}
}

@phdthesis{sauer2024closing,
  title={Closing the Performance Gap Between Multimodal and Public Transit Journey Planning},
  author={Sauer, Jonas},
  year={2024},
  school={Karlsruhe Institute of Technology (KIT)},
  address={Karlsruhe, Germany},
  doi={10.5445/IR/1000173225}
}

@article{pearson1895,
  author    = {Pearson, Karl},
  title     = {Note on Regression and Inheritance in the Case of Two Parents},
  journal   = {Proceedings of the Royal Society of London},
  volume    = {58},
  pages     = {240--242},
  year      = {1895},
  doi       = {10.1098/rspl.1895.0041}
}

@inproceedings{abuaisha2024efficient,
  title={Efficient and exact public transport routing via a transfer connection database},
  author={Abuaisha, Abdallah and Wallace, Mark and Harabor, Daniel and Shen, Bojie},
  booktitle={Proceedings of the International Symposium on Combinatorial Search},
  volume={17},
  pages={2--10},
  year={2024},
  doi = {10.1609/socs.v17i1.31536}
}

@misc{rohovyyi2024ttn,
  author       = {Andrii Rohovyi and Peter J. Stuckey and Toby Walsh},
  title        = {Timetable Nodes for Public Transport Network},
  year         = {2024},
  archivePrefix= {arXiv},
  eprint       = {2410.15715},
  primaryClass = {cs.DS}
}

@inproceedings{rohovyi2025multimodal,
  author    = {Rohovyi, Andrii and Stuckey, Peter J. and Walsh, Toby},
  title     = {Multimodal Pathfinding with Personalized Travel Speed and Transfers of Unlimited Distance},
  booktitle = {Proceedings of the 37th IEEE International Conference on Tools with Artificial Intelligence (ICTAI'25)},
  pages     = {925--931},
  publisher = {IEEE},
  year      = {2025}
}

@inproceedings{phan2019hub,
  author    = {Duc-Minh Phan and Laurent Viennot},
  title     = {Fast Public Transit Routing with Unrestricted Walking Through Hub Labeling},
  booktitle = {Analysis of Experimental Algorithms -- Special Event, {SEA}\textsuperscript{2} 2019},
  series    = {Lecture Notes in Computer Science},
  volume    = {11544},
  pages     = {237--247},
  publisher = {Springer},
  year      = {2019},
  doi       = {10.1007/978-3-030-34029-2_16}
}

@article{kamargianni2016maas,
  author    = {Maria Kamargianni and Weibo Li and Melinda Matyas and Andreas Sch{\"a}fer},
  title     = {A Critical Review of New Mobility Services for Urban Transport},
  journal   = {Transportation Research Procedia},
  volume    = {14},
  pages     = {3294--3303},
  year      = {2016},
  doi       = {10.1016/j.trpro.2016.05.277}
}

@article{jittrapirom2017maas,
	author = {Peraphan Jittrapirom and Valeria Caiati and Anna-Maria Feneri and Shima Ebrahimigharehbaghi and María González and Jishnu Narayan},
	title = {Mobility as a Service: A Critical Review of Definitions, Assessments of Schemes, and Key Challenges},
	journal = {Urban Planning},
	volume = {2},
	number = {2},
	year = {2017},
	issn = {2183-7635},	
    pages = {13--25},
    doi = {10.17645/up.v2i2.931},
	url = {https://www.cogitatiopress.com/urbanplanning/article/view/931}
}

\end{document}